
\font\rmu=cmr10 scaled\magstephalf
\font\bfu=cmbx10 scaled\magstephalf

\font\it=cmti10 scaled \magstephalf
\font\bf=cmbx10 scaled\magstephalf
\rmu

\font\rmus=cmr8
\font\rmuss=cmr6
\font\mait=cmmi10 scaled\magstephalf
\font\maits=cmmi7 scaled\magstephalf
\font\maitss=cmmi7
\font\msyb=cmsy10 scaled\magstephalf
\font\msybs=cmsy8 scaled\magstephalf
\font\msybss=cmsy7
\font\bfus=cmbx7 scaled\magstephalf
\font\bfuss=cmbx7
\font\cmeq=cmex10 scaled\magstephalf

\textfont0=\rmu
\scriptfont0=\rmus
\scriptscriptfont0=\rmuss

\textfont1=\mait
\scriptfont1=\maits
\scriptscriptfont1=\maitss

\textfont2=\msyb
\scriptfont2=\msybs
\scriptscriptfont2=\msybss

\textfont3=\cmeq
\scriptfont3=\cmeq
\scriptscriptfont3=\cmeq

\newfam\bmufam  \textfont\bmufam=\bfu
      \scriptfont\bmufam=\bfus \scriptscriptfont\bmufam=\bfuss

\hsize=15.5cm
\vsize=21cm
\baselineskip=16pt   
\parskip=12pt plus  2pt minus 2pt

\def\one{{\mathchoice {\rm 1\mskip-4mu l} {\rm 1\mskip-4mu l}
{\rm 1\mskip-4.5mu l} {\rm 1\mskip-5mu l}}}
\def\C{{\mathchoice
{\setbox0=\hbox{$\displaystyle\rm C$}\hbox{\hbox to0pt
{\kern0.4\wd0\vrule height0.9\ht0\hss}\box0}}
{\setbox0=\hbox{$\textstyle\rm C$}\hbox{\hbox to0pt
{\kern0.4\wd0\vrule height0.9\ht0\hss}\box0}}
{\setbox0=\hbox{$\scriptstyle\rm C$}\hbox{\hbox to0pt
{\kern0.4\wd0\vrule height0.9\ht0\hss}\box0}}
{\setbox0=\hbox{$\scriptscriptstyle\rm C$}\hbox{\hbox to0pt
{\kern0.4\wd0\vrule height0.9\ht0\hss}\box0}}}}

\font\fivesans=cmss10 at 4.61pt
\font\sevensans=cmss10 at 6.81pt
\font\tensans=cmss10
\newfam\sansfam
\textfont\sansfam=\tensans\scriptfont\sansfam=\sevensans\scriptscriptfont
\sansfam=\fivesans
\def\sans{\fam\sansfam\tensans}
\def\Z{{\mathchoice
{\hbox{$\sans\textstyle Z\kern-0.4em Z$}}
{\hbox{$\sans\textstyle Z\kern-0.4em Z$}}
{\hbox{$\sans\scriptstyle Z\kern-0.3em Z$}}
{\hbox{$\sans\scriptscriptstyle Z\kern-0.2em Z$}}}}

\newcount\foot
\foot=1
\def\note#1{\footnote{${}^{\number\foot}$}{\ftn #1}\advance\foot by 1}

\def\frac#1#2{{#1\over #2}}
\def\text#1{\quad{\hbox{#1}}\quad}

\font\ch=cmbx12 scaled\magstephalf
\font\ftn=cmr8 scaled\magstephalf

\font\it=cmti10 scaled\magstephalf
\font\bf=cmbx10 scaled\magstephalf
\font\titch=cmbx12 scaled\magstep2
\font\titname=cmr10 scaled\magstep2
\font\titit=cmti10 scaled\magstep1
\font\titbf=cmbx10 scaled\magstep2

\nopagenumbers

\line{\hfil AEI-030}
\line{\hfil March 14, 1997}
\vskip2cm
\centerline{\titch Simplifying the spectral analysis of the}
\vskip.5cm
\centerline{\titch volume operator}
\vskip1.5cm
\centerline{{\titname R. Loll}\footnote*{e-mail: loll@aei-potsdam.mpg.de}}
\vskip1cm
\centerline{\titit Max-Planck-Institut f\"ur Gravitationsphysik}
\vskip.2cm
\centerline{\titit Schlaatzweg 1}
\vskip.2cm
\centerline{\titit D-14473 Potsdam, Germany}
\vskip.3cm

\vskip2.2cm
\centerline{\titbf Abstract}
\vskip0.2cm
The volume operator plays a central role in both the kinematics and
dynamics of canonical approaches to quantum gravity which are based on
algebras of generalized Wilson loops. We introduce a method for 
simplifying its spectral analysis, for quantum states that can be
realized on a cubic three-dimensional lattice. This involves a
decomposition of Hilbert space into sectors transforming according to
the irreducible representations of a subgroup of the cubic group.
As an application, we determine the complete spectrum for a class of
states with six-valent intersections.
\vskip0.3cm
\noindent PACS: 04.60.Ds, 04.60.Nc, 02.20.Rt

\noindent keywords: canonical quantum gravity, volume operator, 
lattice gravity, determinant of the metric, octagonal group
\vfill\eject
\footline={\hss\tenrm\folio\hss}
\pageno=1

\line{\ch 1 Introduction\hfil}

Researchers in gravitational physics these days can look back on
a ten-year long effort of quantizing the theory canonically in terms
of a set of ``new'' connection variables [1]. These developments
have led to many new insights, but at the same time have not been free of 
Irrungen und Wirrungen, and the process is by no means finished yet. 

From a technical point of view,    
the problem of representing non-polynomial quantities in the quantum theory
(a central difficulty in ADM-type quantization approaches) in this
formulation has now been recast into that of diagonalizing the operator 
$\hat {\det E}$, where $\det E$ is up to a sign the classical
determinant $\det g$ of the Riemannian metric $g_{ab}$ on spatial slices. 
The fact that non-polynomiality can be re-expressed as ``polynomiality 
modulo powers of $\sqrt{\det g}$'' is not specific to
the connection formulation (see, for example, [2]). However, what 
distinguishes this new approach is the fact
that (certain functions of) $\det g$ have
well-defined finite, self-adjoint operator analogues. 

This happens because in the quantization the classical conjugate
variable pairs $(A_a^{i},E^{a}_{i})$ -- like in Yang-Mills theory --
are represented by multiplication and differentiation with respect
to $A_{a}^{i}$. This implies that suitably quantized versions of 
the classical determinant

$$
\det E(x) =\frac{1}{3!} \eta_{abc}\epsilon^{ijk}
E_{i}^{a}(x)E_{j}^{b}(x)E_{k}^{c}(x)
\eqno(1.1)
$$

\noindent are third-order 
differential operators. Moreover, if the quantization is
based on one-dimensio\-nal flux line states, as is the case in the so-called
``loop quantization'' schemes [3,4], and also in lattice-discretized 
versions of canonical quantum gravity [5], $E$-flux is quantized. This 
is ultimately responsible for the finiteness of quantities like 
the volume operator, 
which is the quantization of the integral 
$\int_{\cal R} d^{3}x\,\sqrt{\det E}$,
for a spatial region $\cal R$.

The framework we will be using in the following is not the one
originally proposed in [1], since the reality
conditions that have to be imposed on its $sl(2,\C)$-valued connections 
$A_a^i(x)$ seem to
be intractable in the quantum theory. Instead, we will use a closely
related version in terms of real, $su(2)$-valued connection forms [6], 
which avoids this difficulty.
In this formulation, the Hamiltonian constraint function is
non-polynomial. 

The subject of this paper is the introduction of a method for simplifying 
the diago\-na\-lization of the volume 
operator (and potentially other operators
relevant in loop quantum gravity), by exploiting symmetries of the
Wilson loop states that form the Hilbert space it is defined on.  
Our discussion will take place in the discretized version of the
theory on a cubic lattice. The discrete volume operator one can 
define on the lattice [7,8] is closely related to the 
finite volume operators one  
obtains after regularization in the continuum theory [9-14].
The type of intersection of Wilson loop states that can occur on the 
lattice is at most six-valent, and therefore not the most general 
from the point of view of the continuum theory, but also in this 
case it may well be sufficiently generic from a physical standpoint.

The operator $\hat {\det E}$ occurs in a variety of contexts. It was 
originally conceived as an ingredient in the definition of the 
volume operator, i.e. the
quantization of $\int_{\cal R} d^{3}x\,\sqrt{\det E}$ 
mentioned above [9,10]. Defined
on the kinematical Hilbert space of the loop representation, this kind
of geometric operator enables one to associate well-known geometric
properties with Wilson loop states, or possibly coarse-grained 
ensembles of such states [9]. 

With the return to {\it real} connection variables in canonical 
quantum gravity, knowledge of the spectrum of $\hat {\det E}$ has
become a necessary prerequisite in the study of the quantum 
Hamiltonian constraint. The fact that the quantum dynamics can still be made
well-defined, in spite of the non-polynomiality of the Hamiltonian,
was first demonstrated in the context of 
the lattice theory [15]. A quantum Hamiltonian for the continuum theory
was constructed in [16]. In order to get rid of the inverse powers 
of $\hat {\det E}$ that occur naturally, an identity was employed by
which the dreibeins $e_{a}^{i}$ (which are non-polynomial functions of
the inverse, densitized dreibeins $E^a_{i}$ that constitute half of 
the basic canonical variables) can be re-expressed as the Poisson brackets
of $A_a^i$ with the total volume function, 
$e_{a}^{i}(x)=2\{ A_{a}^{i}(x), \int d^{3}y\,\sqrt{{\det E}(y)}\}$. 
(This identity holds only if no modulus signs are used around the
${\det E}$-term, otherwise one has to include a factor 
of ${\rm sign}({\det E})$ in the equation.)
In the quantization one substitutes the Poisson brackets by 
commutators. Again the volume operator plays a pivotal role.

Another role of the operator $\hat {\det E}$ was pointed out 
recently [17]. This is related to the fact that one has to impose a
constraint $\det E>0$ classically, if the basic variables are chosen 
to be the Yang-Mills conjugate pairs $(A_a^{i},E^{a}_{i})$. This
constraint distinguishes the phase space of gravity already at the
kinematical level from that of a gauge theory. In the lattice theory,
one can show that all non-vanishing eigenvalues of $\hat {\det E}$ come 
in pairs of opposite sign [17], and a quantum analogue of ${\det E}>0$ can be 
imposed. In practice, this requires an explicit diagonalization of the
operator $\hat {\det E}$.

Finally, another application is the possible inclusion of a 
cosmological constant term of the form 
$\lambda \int d^{3}x\,\sqrt{{\det E}(x)}$ 
in the Hamiltonian, with the integral taken 
over the entire spatial manifold. There are suggestions that
this may be necessary in order to construct a non-trivial continuum
limit of the (discretized) theory [18].

Our present knowledge of the spectrum of $\hat{\det E}$ or the 
volume operator is only partial. It was observed 
in [9,10] that its spectrum is discrete and non-zero contributions 
can only come from intersections of the Wilson loop states it acts on. 
Its diagonalization can in a so-called spin-network basis [19] be reduced
to a diagonalization on finite-dimensional subspaces of Hilbert space.
In [7], we gave a general proof for why intersections have to be at
least four-valent in order to give a non-trivial contribution. 
Part of the non-vanishing spectrum for such four-valent intersections
was first given in [8]. These calculations were confirmed in [11], where 
also a general formula for the matrix elements of the volume operator
was derived. These were given in terms of an orthogonal basis, obtained 
by decomposing spin-network states (closely related to Wilson loop
states) with $n$-valent intersections into three-valent ones.
Another general expression for matrix elements, with respect to a 
different orthogonal basis, was given in [14], together with  
formulae for spectra for simple special cases of classes of four-valent 
intersections. (In order to avoid confusion, it should be pointed out
that different volume operators may differ by overall factors and
modulus signs (see, for example, the comments in [13]). 
Still, results on their spectra tend to be closely related.)

As we mentioned above, $\hat {\det E}$ can be diagonalized on 
finite-dimensional subspaces. These are given by fixing the flux line
(or spin) assignments on the edges incoming at a given intersection 
where $\hat {\det E}$ acts. Another way of saying this is that there
exist operators associated with these edges that commute with 
$\hat {\det E}$. On the lattice, these are of the form 
$\sqrt{\sum_{i}\hat E^{a}_{i}(l)\hat E^{ai}(l)}$ (no sum over $a$) with spectrum
$\sqrt{ j(j+2)/4}$, $j=0,1,2,\ldots$, and related to the measurement 
of area in the $a$-direction [20]. 

One does not expect to be able to derive analytic formulae for the spectrum 
of $\hat {\det E}$ as the dimension of the finite-dimensional 
subspaces obtained by fixing the quantum numbers $j$ for the 
incoming edges grows. The explicit evaluation of the spectrum is 
therefore limited by one's computational ability to diagonalize large matrices. 
A suitable choice of basis in which the matrix elements are evaluated will
eventually also be determined by the form of the Hamiltonian.

In this paper we construct further local operators, related to the geometry 
of the intersection of a spin network state, that commute with 
$\hat {\det E}$, and lead to a further reduction of the 
finite-dimensional Hilbert spaces into smaller subspaces on which 
$\hat {\det E}$ can be diagonalized separately. 
In mathematical terms, we will be analyzing the irreducible
representations of a discrete local symmetry group of the quantum
states, and decompose the Hilbert space accordingly. The method is
general and can also be applied to other operators, most importantly
the Hamiltonian constraint. This leads to a considerable 
simplification, as we will demonstrate for a class of six-valent 
intersections, with some of the resulting ``superselection sectors'' 
becoming totally non-degenerate with respect to their volume spectrum.

Our analysis will take place on a cubic lattice, and is therefore valid
for all quantum states that can be realized on connected subsets of
edges from such a lattice. The discrete symmetry group is in this case 
the 24-element group ${\cal O}$, the so-called cubic or orthogonal
group [21]. The analysis may be generalized to intersections with
a different geometry and symmetry group. 

In the next section, we introduce a labelling of local spin network
states (more precisely, a labelling for the so-called intertwiners
at an intersection) with a simple transformation behaviour under the 
cubic group. For the construction of the superselection sectors, the
relevant discrete group is a subgroup ${\cal O}^{(6)}$ of the cubic 
group times the $\Z_2$-factor associated with total spatial reflection. 
We describe its irreducible representations and give an explicit
construction for states transforming according to a given irreducible 
representation from the elements of 
${\cal O}^{(6)}\times\Z_{2}$-orbits. In Section 3 
we give the explicit form of the operator $\hat {\det E}$ associated 
with a vertex (or intersection) $n$. For a special class of six-valent 
intersections, corresponding to flux line assignments 
$\vec  j=(j,j,j,j,j,j)$, we perform the orbit decomposition, the 
construction of the superselection sectors, and diagonalization of
$\hat {\det E}$ up to $j=10$. All eigenvalues, together with their
multiplicities in the individual sectors, are listed in Table 9.
We end in Section 4 with a summary of our results and a brief
discussion of the condition ${\det E}>0$. 

\vskip1.5cm
\line{\ch 2 Action of the orthogonal group on the quantum state space\hfil}

Our kinematical quantum state space is the gauge-invariant sector of 
the Hilbert space of an $SU(2)$-lattice gauge theory in the Hamiltonian
formulation [22]. Following the philosophy of spin network states [19], we
use flux line labels $j=0,1,2,\ldots$ for lattice edges or links $l$. As an 
overcomplete set of labels for the contractors or intertwiners 
situated at a vertex $n$ of the lattice we use 9 numbers $j_{ik}$,
$i,k=1,2,3$ [17]. The index $i$ denotes the incoming edges at $n$ from the
1-, 2-, or 3-direction (with respect to some fixed orientation for
the three lattice axes), and the index $k$ the 
corresponding outgoing edges. That is, $j_{12}$ denotes the number of
(unoriented) flux lines routed through $n$ between the (-1)- and the
2-direction, etc. It is convenient to arrange the set of nine 
numbers into matrix form,   

$$
J:=
\left(\matrix{j_{11}&j_{12}&j_{13}\cr j_{21}&j_{22}&j_{23}\cr
j_{31}&j_{32}&j_{33}\cr}\right).
\eqno(2.1)
$$

Given flux line assignments $j_i$, $i=-1,-2,-3,1,2,3$, 
for the in- and outgoing links, 
it is easy to generate all allowed intertwiner configurations $J$. 
The elements of the rows and columns
of $J$ simply have to add up to the appropriate $j_i$, for example,
$\sum_{i=1}^3 j_{1,i}=j_{-1}$, $\sum_{i=1}^3 j_{i,1}=j_{1}$.
General elements of the cubic group $\cal O$ do not map intertwiners 
of the form (2.1) into themselves (but lead to configurations that
by virtue of Mandelstam identities can be re-expressed as linear 
combinations of $J$-configurations). However, a six-dimensional 
subgroup ${\cal O}^{(6)}$ of $\cal O$ leaves the label set
invariant. 
Apart from the identity transformation, the 
non-trivial elements of this subgroup 
${\cal O}^{(6)}$ act on the $J$'s according to 

$$
\eqalign{
&R_1(J):=\left(\matrix{j_{11}&j_{31}&j_{21}\cr j_{13}&j_{33}&j_{23}\cr
  j_{12}&j_{32}&j_{22}\cr}\right),\;
R_2(J):=\left(\matrix{j_{33}&j_{23}&j_{13}\cr j_{32}&j_{22}&j_{12}\cr
  j_{31}&j_{21}&j_{11}\cr}\right),\;
R_3(J):=\left(\matrix{j_{22}&j_{12}&j_{32}\cr j_{21}&j_{11}&j_{31}\cr
  j_{23}&j_{13}&j_{33}\cr}\right),\cr
&\hskip2cm S_1(J):=\left(\matrix{j_{22}&j_{23}&j_{21}\cr 
  j_{32}&j_{33}&j_{31}\cr
  j_{12}&j_{13}&j_{11}\cr}\right),\;
S_2(J):=\left(\matrix{j_{33}&j_{31}&j_{32}\cr j_{13}&j_{11}&j_{12}\cr
  j_{23}&j_{21}&j_{22}\cr}\right).}
\eqno(2.2)
$$

\vskip.7cm
{\offinterlineskip\tabskip=0pt
 \halign{ \strut\vrule#& \quad # \quad &\vrule#&
          \quad\hfil #\quad &
          \quad\hfil #\quad &
          \quad\hfil #\quad &
          \quad\hfil #\quad &
          \quad\hfil #\quad &
          \quad\hfil #\quad &\vrule#\cr \noalign{\hrule}
 & & & & & & && &\cr
 & && $\one $ & $R_1$ & $R_2$
              & $R_3$ & $S_1$ & $S_2$ &\cr
 & & & & & & && &\cr
 \noalign{\hrule}
 & & & & & & && &\cr
 & $\one$ &&
      $\one$ & $R_1$ & $R_2$ & $R_3$ & $S_1$ & $S_2$ &\cr
 & & & & & & && &\cr
 & $R_1$ &&
      $R_1$ & $\one$ & $S_1$ & $S_2$ & $R_2$ & $R_3$ &\cr
 & & & & & & && &\cr
 & $R_2$ &&
      $R_2$ & $S_2$ & $\one$ & $S_1$ & $R_3$ & $R_1$ &\cr
 & & & & & & && &\cr
 & $R_3$ &&
      $R_3$ & $S_1$ & $S_2$ & $\one$ & $R_1$ & $R_2$ &\cr
 & & & & & & && &\cr
 & $S_1$ &&
      $S_1$ & $R_3$ & $R_1$ & $R_2$ & $S_2$ & $\one$ &\cr
 & & & & & & && &\cr
 & $S_2$ &&
      $S_2$ & $R_2$ & $R_3$ & $R_1$ & $\one$ & $S_1$ &\cr
 & & & & & & && &\cr
 \noalign{\hrule} }
\normalbaselines
\baselineskip=16pt   
\vskip.5cm
\line{{\bf Table 1} \hskip.3cm Multiplication table for the subgroup 
${\cal O}^{(6)}$ of the octagonal group.\hfill}
\vskip.7cm

\noindent The multiplication table for the group ${\cal O}^{(6)}$ is
given in Table 1. We will also need the total space reflection $T$, 
corresponding to the transform of the matrix $J$,

$$
T(J):=\left(\matrix{j_{11}&j_{21}&j_{31}\cr j_{12}&j_{22}&j_{32}\cr
  j_{13}&j_{23}&j_{33}\cr}\right).
\eqno(2.3)
$$

\noindent Since $T$ commutes with all elements of ${\cal O}^{(6)}$,
adjoining it we obtain a 12-element group 
${\cal O}^{(6)} \times T \equiv {\cal O}^{(6)} \times\Z_2$. This 
group is important because it is a subgroup of 
the classical invariance group of the 
lattice function $({\det E})^{2}$ (see Section 3 below). 
There is still a redundancy in the
set of allowed $J$'s which is associated with Mandelstam constraints 
and consists of all identities of the form

$$
\eqalign{
&\left(\matrix{j_{11}+1&j_{12}&j_{13}\cr j_{21}&j_{22}+1&j_{23}\cr
  j_{31}&j_{32}&j_{33}+1\cr}\right)
-\left(\matrix{j_{11}+1&j_{12}&j_{13}\cr j_{21}&j_{22}&j_{23}+1\cr
  j_{31}&j_{32}+1&j_{33}\cr}\right)\cr
-&\left(\matrix{j_{11}&j_{12}+1&j_{13}\cr j_{21}+1&j_{22}&j_{23}\cr
  j_{31}&j_{32}&j_{33}+1\cr}\right)+
\left(\matrix{j_{11}&j_{12}+1&j_{13}\cr j_{21}&j_{22}&j_{23}+1\cr
  j_{31}+1&j_{32}&j_{33}\cr}\right)\cr
+&\left(\matrix{j_{11}&j_{12}&j_{13}+1\cr j_{21}+1&j_{22}&j_{23}\cr
  j_{31}&j_{32}+1&j_{33}\cr}\right)
-\left(\matrix{j_{11}&j_{12}&j_{13}+1\cr j_{21}&j_{22}+1&j_{23}\cr
  j_{31}+1&j_{32}&j_{33}\cr}\right)=0.}\eqno(2.4)
$$

As already mentioned in the introduction, 
our aim is to identify the irreducible representations
of the discrete group ${\cal O}^{(6)}$ (${\cal O}^{(6)}\times \Z_2$), 
and construct
corresponding superselection sectors on which the operator $\hat {\det 
E}$ ($(\hat {\det E})^{2}$) can be diagonalized separately. 
The group ${\cal O}^{(6)}$ contains three conjugacy classes of elements
namely, $\{\one\},\{R_1,R_2,R_3\}$ and $\{S_1,S_2\}$. 

\vskip1cm
{\offinterlineskip\tabskip=0pt
 \halign{ \strut\vrule#& \quad # \quad &\vrule#&
          \quad\hfil #\quad &
          \quad\hfil #\quad &
          \quad\hfil #\quad &
          \quad\hfil #\quad &
          \quad\hfil #\quad &
          \quad\hfil #\quad &\vrule#\cr \noalign{\hrule}
 & & & & & & && &\cr
 & && $\{\one\} $ & $\{ R_i\}$ & $\{S_i\}$
              & $\{ T\}$ & $\{ TR_i\}$ & $\{TS_i\}$ &\cr
 & & & & & & && &\cr
 \noalign{\hrule}
 & & & & & & && &\cr
 & $A_1^+$ &&
      $1$ & $1$ & $1$ & $1$ & $1$ & $1$ &\cr
 & & & & & & && &\cr
 & $A_2^+$ &&
      $1$ & $-1$ & $1$ & $1$ & $-1$ & $1$ &\cr
 & & & & & & && &\cr
 & $E^+$ &&
      $2$ & $0$ & $-1$ & $2$ & $0$ & $-1$ &\cr
 & & & & & & && &\cr
 & $A_1^-$ &&
      $1$ & $1$ & $1$ & $-1$ & $-1$ & $-1$ &\cr
 & & & & & & && &\cr
 & $A_2^-$ &&
      $1$ & $-1$ & $1$ & $-1$ & $1$ & $-1$ &\cr
 & & & & & & && &\cr
 & $E^-$ &&
      $2$ & $0$ & $-1$ & $-2$ & $0$ & $1$ &\cr
 & & & & & & && &\cr
 \noalign{\hrule} }
\normalbaselines
\baselineskip=16pt   
\vskip.5cm
\line{{\bf Table 2} \hskip.3cm Character table for the group  
${\cal O}^{(6)}\times \Z_2$.\hfill}
\vskip.7cm

\noindent Following [21],
one establishes the existence of three irreducible representations: two 
one-dimensional ones (called $A_1$ and $A_2$) and one two-dimensional one
(called $E$). They can be identified by the values of their characters,
i.e. the traces of the matrices representing the group elements (which
only depend on the conjugacy class). The enlarged group
${\cal O}^{(6)}\times T$ has six conjugacy classes and six irreducible
representations, since each of the previous re\-pre\-sentations gives rise
to one of positive and one of negative parity, denoted by a subscript
$+$ or $-$.
The characters for the group ${\cal O}^{(6)}\times T$ are given in 
Table 2. 

In order to establish the contents of irreducible representations of
some general representation, one can make use of the following
character formula. It relates the multiplicity $m_R$ of a given
irreducible representation $R$ in a general representation $\cal M$
to the number $n_K$ of elements in the conjugacy class $K$, the
traces $\chi^{\cal M}_K$ of matrices in the representation $\cal M$
in class $K$, and the characters $\chi_K^R$ of the irreducible representation
$R$,

$$
m_R=\frac{1}{d}\sum_K n_K\, \chi_K^{\cal M}\, \chi_K^R.
\eqno(2.5)
$$

\noindent For the group ${\cal O}^{(6)}$, $d=6$, and for the group
${\cal O}^{(6)}\times \Z_2$, $d=12$.

The possible orbit sizes that occur under the action of the group
${\cal O}^{(6)}\times \Z_2$ on states of the form (2.1) are 1, 2, 3,
6 and 12. We distinguish between parity-even and parity-odd orbits.
In the former, all elements are parity-invariant, i.e. $T(J)=J$,
whereas in the latter $T(J)\not=J$ for all $J$. The contents of
irreducible representations of the individual orbit types is listed
in Table 3 (the numbers in brackets denote the number of $J$-states).

\vskip.7cm
{\offinterlineskip\tabskip=0pt
 \halign{ \strut\vrule#& \quad # \quad &\vrule#&
          \quad #\hfil \quad &\vrule#\cr \noalign{\hrule}
 & && &\cr
 & 1-d even && $A_1^+\, (1)$ &\cr
 & && &\cr
 & 2-d odd && $A_1^+\, (1)$, $A_1^-\, (1)$ & \cr
 & && &\cr
 & 3-d even && $A_1^+\, (1)$, $E^+\, (2)$ & \cr
 & && &\cr
 & 6-d odd && $A_1^+\, (1)$, $A_1^-\, (1)$, $E^+\, (2)$, $E^-\, (2)$ & \cr
 & && &\cr
 & 6-d even && $A_1^+\, (1)$, $A_2^+\, (1)$, $E^+\, (4)$ & \cr
 & && &\cr
 & 12-d odd && $A_1^+\, (1)$, $A_1^-\, (1)$, $A_2^+\, (1)$, $A_2^-\, (1)$,
     $E^+\, (4)$, $E^-\, (4)$ & \cr
 & && &\cr
 \noalign{\hrule} }
\normalbaselines
\baselineskip=16pt   
\vskip.5cm
\line{{\bf Table 3} \hskip.3cm Representation contents of the 
${\cal O}^{(6)}\times\Z_2$-orbits.\hfill}
\vskip.7cm

Instead of diagonalizing a maximal set of commuting operators built 
from the elements of ${\cal O}^{(6)}\times\Z_2$ (for example,
given by $\{\one, R_{1}+R_{2}+R_{3},S_{1}+S_{2},T\}$) in a given 
finite-dimensional sector of Hilbert space, we have found it
computationally simpler to first construct the orbits and from those 
the states transforming according to a definite irreducible
representation. One way of how this may be done is illustrated in 
Table 4 (by $J$ we denote in this context an arbitrary, fixed element 
of an orbit). 
Note that $A_2^+$-states change sign whenever one of the $R_i$ is applied. 
The prescriptions for the $E$-states in the two-dimensional 
representations are of course non-unique.
Remember also that (2.4) induces a residual redundancy in the sets of 
states constructed according to Table 4, leading to relations among
elements from different orbits.
For the explicit calculations of the next section 
(which were performed using Mathematica on a DEC AlphaStation 255 4/232
with 64 Mb RAM) these could be handled without particular problems.

\vskip.7cm
{\offinterlineskip\tabskip=0pt
 \halign{ \strut\vrule#& \quad # \quad &\vrule#&
          \quad #\hfil \quad &\vrule#\cr \noalign{\hrule}
 & && &\cr
 & $A_{1}^{+}$ && all orbits: 
      $(\one +T)(J+R_{1}J+R_{2}J+R_{3}J+S_{1}J+S_{2}J)$ &\cr
 & && &\cr
 & $A_{1}^{-}$ && all odd orbits: 
      $(\one -T)(J+R_{1}J+R_{2}J+R_{3}J+S_{1}J+S_{2}J)$ & \cr
 & && &\cr
 & $A_{2}^{+}$ && 6-d even orbit: $J+S_1J+S_2J-R_1J-R_1S_1J-R_1S_2J$ & \cr
 & && 12-d odd orbit: $(\one+T)(J+S_1J+S_2J-R_1J-R_1S_1J-R_1S_2J)$ & \cr
 & && &\cr
 & $A_{2}^{-}$ && 12-d odd orbit: 
      $(\one+T)(J+S_1J+S_2J-R_1J-R_1S_1J-R_1S_2J)$ & \cr
 & && &\cr
 & $E^{+}$ && 3-d even orbit: $J-S_1J$, $J-S_2J$, & \cr
 & && 6-d odd orbit: $(\one +T)(J-S_1J)$, $(\one +T)(J-S_2J)$ & \cr
 & && 6-d even orbit: $(\one-S_1)(J+R_1J)$, $(\one-S_2)(J+R_1J)$, 
    $(\one-S_1)(R_2J+S_2J)$, &\cr
 & && \hskip3cm $(\one-S_2)(R_2J+S_2J)$ & \cr 
 & && 12-d odd orbit: 
    $(\one +T)(\one-S_1)(J+R_1J)$, $(\one +T)(\one-S_2)(J+R_1J)$, & \cr
 & && \hskip3cm 
    $(\one +T)(\one-S_1)(R_2J+S_2J)$, $(\one +T)(\one-S_2)(R_2J+S_2J)$ & \cr 
 & && &\cr
 & $E^{-}$ && 6-d odd orbit: $(\one -T)(J-S_1J)$, $(\one +T)(J-S_2J)$ & \cr
 & && 12-d odd orbit: 
    $(\one -T)(\one-S_1)(J+R_1J)$, $(\one -T)(\one-S_2)(J+R_1J)$, & \cr 
 & && \hskip3cm 
    $(\one -T)(\one-S_1)(R_2J+S_2J)$, $(\one -T)(\one-S_2)(R_2J+S_2J)$ & \cr 
 & && &\cr
 \noalign{\hrule} }
\normalbaselines
\baselineskip=16pt   
\vskip.5cm
\line{{\bf Table 4} \hskip.3cm How to construct states transforming in
a given irreducible representation from the\hfill}
\line{\hskip1.8cm elements of 
${\cal O}^{(6)}\times\Z_2$-orbits.\hfill}
\vskip.7cm
\vfill\eject
 
\line{\ch 3 Determining the spectrum of the operator (det E)\hfil}

We will now apply the results of the previous section to the 
discretized version ${\det E}(n)$ of the function (1.1), acting at a 
lattice vertex $n$. The 
corresponding self-adjoint quantum operator we will call $\hat D(n)$ 
(for convenience rescaled by a factor of $\frac{1}{6}$
with respect to the definition in [8]). 
In terms of the symmetrized link momenta $\hat p_i(n,\hat a)$, it is 
given as

$$
\hat D(n):=\frac{1}{3!} \eta_{\hat a \hat b \hat c}\epsilon^{ijk}
\hat p_{i}(n,\hat a) \hat p_{j}(n,\hat b) \hat p_{k}(n,\hat c),
\eqno(3.1)
$$
\noindent where
$$
\hat p_i(n,\hat a)=\frac{i}{2}(X_+^i(n,\hat a)+X_-^i(n-1_{\hat a},\hat a)),
\eqno(3.2)
$$

\noindent and $X_\pm^i(n,\hat a)$ denote the left- and right-invariant
vector fields on the group manifold associated with the link $l=(n,\hat a)$,
with commutators $[X_{\pm}^{i},X_{\pm}^{j}]=\pm\epsilon^{ijk}X_{\pm}^{k}$.

The key observation is that the classical lattice function $D(n)\equiv
{\det E}(n)$ is invariant under the action of the 24-element cubic group 
$\cal O$ [21] (whose elements can be thought
of as the orientation-preserving permutations
of the three (oriented) lattice axes meeting at the
intersection $n$). By contrast, $D(n)$ changes sign under the total
space reflection $T$ (i.e. under simultaneous inversion of the three axes).  

As a result of this classical symmetry, 
eigenstates of $\hat D(n)$ can be classified according
to the irreducible representations of $\cal O$. 
This set-up is familiar to lattice gauge theorists,
because it has been employed in analyzing
the glueball spectrum of the Hamiltonian in four-dimensional
$SU(3)$-lattice gauge theory [23]. As explained in Section 2, in 
the present $SU(2)$-context it is convenient to work with a set of
local states that is partially gauge-fixed with respect to the
$\cal O$-action, leaving us with 
${\cal O}^{(6)}$ as the residual symmetry group.

The action of the operator $\hat D(n)$ on states of the form (2.1)
can be obtained by considering all possible ways in which the
triple derivatives contained in $\hat D(n)$ can act on sets of
flux lines routed through the vertex $n$. Its explicit form is
given by 

$$
\hat D\left(\matrix{j_{11}&j_{12}&j_{13}\cr j_{21}&j_{22}&j_{23}\cr
j_{31}&j_{32}&j_{33}\cr}\right) =
$$

$$
\eqalign{
&\frac{i}{4} j_{11}j_{22}j_{33}\left[
\left(\matrix{j_{11}-1&j_{12}+1&j_{13}\cr j_{21}&j_{22}-1&j_{23}+1\cr
j_{31}+1&j_{32}&j_{33}-1\cr}\right)-
\left(\matrix{j_{11}-1&j_{12}&j_{13}+1\cr j_{21}+1&j_{22}-1&j_{23}\cr
j_{31}&j_{32}+1&j_{33}-1\cr}\right)\right]\cr
+&\frac{i}{16}j_{12}j_{23}j_{31}\left[
\left(\matrix{j_{11}&j_{12}-1&j_{13}+1\cr j_{21}+1&j_{22}&j_{23}-1\cr
j_{31}-1&j_{32}+1&j_{33}\cr}\right)-
\left(\matrix{j_{11}+1&j_{12}-1&j_{13}\cr j_{21}&j_{22}+1&j_{23}-1\cr
j_{31}-1&j_{32}&j_{33}+1\cr}\right)\right]\cr
+&\frac{i}{16}j_{13}j_{21}j_{32}\left[
\left(\matrix{j_{11}+1&j_{12}&j_{13}-1\cr j_{21}-1&j_{22}+1&j_{23}\cr
j_{31}&j_{32}-1&j_{33}+1\cr}\right)-
\left(\matrix{j_{11}&j_{12}+1&j_{13}-1\cr j_{21}-1&j_{22}&j_{23}+1\cr
j_{31}+1&j_{32}-1&j_{33}\cr}\right)\right]\cr
+&\frac{i}{16}j_{11}j_{32}(j_{12}+2 j_{33}-j_{13}-j_{21}+
2 j_{22}+j_{31}+2)
\left(\matrix{j_{11}-1&j_{12}+1&j_{13}\cr j_{21}&j_{22}&j_{23}\cr
j_{31}+1&j_{32}-1&j_{33}\cr}\right)\cr
+&\frac{i}{16}j_{11}j_{23}(j_{12}-2 j_{22}-j_{13}-j_{21}
-2 j_{33}+j_{31}-2)
\left(\matrix{j_{11}-1&j_{12}&j_{13}+1\cr j_{21}+1&j_{22}&j_{23}-1\cr
j_{31}&j_{32}&j_{33}\cr}\right)\cr
+&\frac{i}{16}j_{22}j_{13}(j_{23}+2 j_{11}-j_{21}-j_{32}+
2 j_{33}+j_{12}+2)
\left(\matrix{j_{11}&j_{12}+1&j_{13}-1\cr j_{21}&j_{22}-1&j_{23}+1\cr
j_{31}&j_{32}&j_{33}\cr}\right)\cr
+&\frac{i}{16}j_{22}j_{31}(j_{23}-2 j_{33}-j_{21}-
j_{32}-2 j_{11}+j_{12}-2)
\left(\matrix{j_{11}&j_{12}&j_{13}\cr j_{21}+1&j_{22}-1&j_{23}\cr
j_{31}-1&j_{32}+1&j_{33}\cr}\right)\cr
+&\frac{i}{16}j_{33}j_{21}(j_{31}+2 j_{22}-j_{32}-j_{13}+
2 j_{11}+j_{23}+2)
\left(\matrix{j_{11}&j_{12}&j_{13}\cr j_{21}-1&j_{22}&j_{23}+1\cr
j_{31}+1&j_{32}&j_{33}-1\cr}\right)\cr
+&\frac{i}{16}j_{33}j_{12}(j_{31}-2 j_{11}-j_{32}-
j_{13}-2 j_{22}+j_{23}-2)
\left(\matrix{j_{11}&j_{12}-1&j_{13}+1\cr j_{21}&j_{22}&j_{23}\cr
j_{31}&j_{32}+1&j_{33}-1\cr}\right)\cr
-&\frac{i}{16}j_{23}(j_{11}j_{12}+j_{12}j_{22}+j_{12}j_{33}
+j_{12}j_{13}+2 j_{12})
\left(\matrix{j_{11}&j_{12}-1&j_{13}+1\cr j_{21}&j_{22}+1&j_{23}-1\cr
j_{31}&j_{32}&j_{33}\cr}\right)\cr
-&\frac{i}{16}j_{31}(j_{22}j_{23}+j_{23}j_{33}+j_{11}j_{23}+
j_{21}j_{23}+2 j_{23})
\left(\matrix{j_{11}&j_{12}&j_{13}\cr j_{21}+1&j_{22}&j_{23}-1\cr
j_{31}-1&j_{32}&j_{33}+1\cr}\right)\cr
-&\frac{i}{16}j_{12}(j_{31}j_{33}+j_{11}j_{31}+j_{22}j_{31}+
j_{31}j_{32}+2 j_{31})
\left(\matrix{j_{11}+1&j_{12}-1&j_{13}\cr j_{21}&j_{22}&j_{23}\cr
j_{31}-1&j_{32}+1&j_{33}\cr}\right)\cr
+&\frac{i}{16}j_{13}(j_{11}j_{21}+j_{21}j_{22}+j_{21}j_{33}+
j_{21}j_{23}+2 j_{21})
\left(\matrix{j_{11}+1&j_{12}&j_{13}-1\cr j_{21}-1&j_{22}&j_{23}+1\cr
j_{31}&j_{32}&j_{33}\cr}\right)\cr
+&\frac{i}{16}j_{21}(j_{22}j_{32}+j_{32}j_{33}+j_{11}j_{32}+
j_{31}j_{32}+2 j_{32})
\left(\matrix{j_{11}&j_{12}&j_{13}\cr j_{21}-1&j_{22}+1&j_{23}\cr
j_{31}+1&j_{32}-1&j_{33}\cr}\right)\cr
+&\frac{i}{16}j_{32}(j_{13}j_{33}+j_{11}j_{13}+j_{13}j_{22}+
j_{12}j_{13}+2 j_{13})
\left(\matrix{j_{11}&j_{12}+1&j_{13}-1\cr j_{21}&j_{22}&j_{23}\cr
j_{31}&j_{32}-1&j_{33}+1\cr}\right)}
$$
$$
\eqalign{
+&\frac{i}{8}j_{11}j_{22}(j_{23}-j_{32}+j_{31}-j_{13})
\left(\matrix{j_{11}-1&j_{12}+1&j_{13}\cr j_{21}+1&j_{22}-1&j_{23}\cr
j_{31}&j_{32}&j_{33}\cr}\right)\cr
+&\frac{i}{8}j_{11}j_{33}(j_{12}-j_{21}+j_{23}-j_{32})
\left(\matrix{j_{11}-1&j_{12}&j_{13}+1\cr j_{21}&j_{22}&j_{23}\cr
j_{31}+1&j_{32}&j_{33}-1\cr}\right)\cr
+&\frac{i}{8}j_{22}j_{33}(j_{12}-j_{21}+j_{31}-j_{13})
\left(\matrix{j_{11}&j_{12}&j_{13}\cr j_{21}&j_{22}-1&j_{23}+1\cr
j_{31}&j_{32}+1&j_{33}-1\cr}\right)\cr
+&\frac{i}{16}(\ j_{11}(j_{12}j_{31}-j_{12}j_{32}+j_{13}j_{23}-j_{13}j_{21}+
j_{21}j_{23}-j_{31}j_{32}+j_{23}-j_{32})+\cr
&\hskip.7cm j_{22}(j_{21}j_{31}-j_{12}j_{13}+j_{12}j_{23}-j_{21}j_{32}+
j_{31}j_{32}-j_{13}j_{23}+j_{31}-j_{13})+\cr
&\hskip.7cm j_{33}(j_{12}j_{13}-j_{21}j_{23}+j_{23}j_{31}-j_{21}j_{31}+
j_{12}j_{32}- j_{13}j_{32}+j_{12}-j_{21})+\cr
&j_{12}j_{13}j_{23}-j_{12}j_{13}j_{32}+j_{21}j_{23}j_{31}-j_{13}j_{21}j_{23}+
j_{12}j_{31}j_{32}-j_{21}j_{31}j_{32}+\cr
&j_{12}j_{23}+j_{12}j_{31}-j_{13}j_{21}-j_{13}j_{32}+j_{23}j_{31}-
j_{21}j_{32}\ )
\left(\matrix{j_{11}&j_{12}&j_{13}\cr j_{21}&j_{22}&j_{23}\cr
j_{31}&j_{32}&j_{33}\cr}\right)
}\eqno(3.3)
$$

Note that the $J$-states are not orthogonal with respect to the Haar measure  
on the lattice. Still, we will see that within the superselection sectors
of the cubic group, degeneracy of the eigenvalues of $\hat D(n)$ is largely 
lifted, which makes orthogonality largely automatic. 
One can verify the following conjugation relations by using the 
explicit form (3.3) for the operator $\hat D$:

$$
R_i\hat D R_i=\hat D,\; i=1,2,3,\;\;
S_i\hat D S_i=\ \hat D,\; i=1,2,\;\;
T\hat D T=-\hat D,
\eqno(3.4)
$$

\noindent whence it follows that $\hat D$ obeys the (anti-)commutation relations

$$
[\hat D,R_i]=0,\;i=1,2,3,\;\; 
[\hat D,S_1+S_2]=0,\;
[\hat D,T]_+=0. 
\eqno(3.5)
$$

We conclude that $\hat D$ does not alter the ${\cal O}^{(6)}$-quantum
numbers, but maps positive- into negative-parity states and vice versa. 
This latter property suggests a different approach to the 
diagonalization of $\hat D(n)$. Firstly, it follows from $[\hat D,T]_+=0$ that
$[\hat D^{2},T]=0$. Secondly, we have shown in [17] that eigenstates of 
$\hat D$ can be obtained from those of $\hat D^{2}$. 
For any eigenstate $\chi$ of $\hat D^2$ with eigenvalue $v\not=0$,
$\hat D^2(n)\chi=v^2\chi$, one may construct a pair of eigenstates of 
$\hat D(n)$ with eigenvalues $\pm v$, namely,  
 
$$
\hat D(\chi\pm\frac{1}{|v|}\hat D \chi)=\pm\frac{1}{|v|}\hat D^2\chi+
\hat D\chi=\pm |v| (\chi \pm \frac{1}{|v|}\hat D\chi).
\eqno(3.6)
$$

Since in our search for eigenstates we would like to reduce the size of 
matrices that have to be diagonalized, it is simpler to
diagonalize $\hat D(n)^{2}$ first, and then construct eigenstates
of $\hat D(n)$ using (3.6). The explicit form for $\hat D(n)^2$ contains
142 terms and can be obtained from (3.3).

One further observation turns out to be useful. Since
parity-odd wave functions are constructed by weighted sums (with
factors $\pm1$) of spin network states, which may sometimes vanish,
there are always fewer states transforming according to the representations
$A_i^-$, $E^-$, than those transforming according to $A_i^+$, $E^+$.
The most efficient way of diagonalizing $\hat D$ (and the one which 
we follow below) is therefore to start
from the wave functions transfor\-ming in one of the
negative-parity irreducible representations, diagonalize $\hat D^2$, 
construct the images under $\hat D$ of the resulting states
(which all have positive parity), and then form linear
combinations according to (3.6). The number of zero-volume states is given
by the difference of positive- and negative-parity states.

\vskip1cm
{\offinterlineskip\tabskip=0pt
 \halign{ \strut\vrule#& \quad # \quad &\vrule#&
          \quad\hfil #\quad &
          \quad\hfil #\quad &
          \quad\hfil #\quad &
          \quad\hfil #\quad &
          \quad\hfil #\quad &
          \quad\hfil #\quad &
          \quad\hfil #\quad &
          \quad\hfil #\quad &
          \quad\hfil #\quad &
          \quad\hfil #\quad &\vrule#\cr \noalign{\hrule}
 & & & & & & & & & & && &\cr
 & \hfil j\hfil && 1 & 2 & 3 & 4 & 5 & 6 & 7 & 8 & 9 & 10 &\cr
 & & & & & & & & & & && &\cr
 \noalign{\hrule}
 & & & & & & & & & & && &\cr
 & 1-d even &&
      1 & 2 & 2 & 3 & 3 & 4 & 4 & 5 & 5 & 6 &\cr
 & & & & & & & & & & && &\cr
 & 2-d odd &&
      1 & 2 & 4 & 6 & 9 & 12 & 16 & 20 & 25 & 30 &\cr
 & & & & & & & & & & && &\cr
 & 3-d even &&
      1 & 3 & 5 & 9 & 12 & 18 & 22 & 30 & 35 & 45 &\cr
 & & & & & & & & & & && &\cr
 & 6-d odd &&
      0 & 1 & 4 & 9 & 18 & 30 & 48 & 70 & 100 & 135 &\cr
 & & & & & & & & & & && &\cr
 & 6-d even &&
      0 & 0 & 1 & 2 & 5 & 8 & 14 & 20 & 30 & 40 &\cr
 & & & & & & & & & & && &\cr
 & 12-d odd &&
      0 & 0 & 0 & 1 & 3 & 8 & 16 & 30 & 50 & 80 &\cr
 & & & & & & & & & & && &\cr
 & total &&
      3 & 8 & 16 & 30 & 50 & 80 & 120 & 175 & 245 & 336 &\cr
 & & & & & & & & & & && &\cr
 \noalign{\hrule} }
\normalbaselines
\baselineskip=16pt   
\vskip.5cm
\line{{\bf Table 5} \hskip.3cm Numbers of ${\cal O}^{(6)}\times \Z_2$-orbits 
for states with $\vec j=(j,j,j,j,j,j)$.\hfill}
\vskip.7cm

As an application of the scheme outlined above, we have analyzed the irreducible
representation contents of the sub-Hilbert spaces with 
flux line numbers $\vec j=(j_{-1},j_{-2},j_{-3},j_1,j_2,j_3)=
(j,j,j,j,j,j)$ (i.e. for genuine six-valent intersections) up to $j=10$, and 
obtained the spectrum of $\hat D(n)$. 
This class of configurations is special in the sense 
that states with fixed $j$ are mapped
into themselves by the action of ${\cal O}^{(6)}\times \Z_2$.
The numbers of orbits of a given type for fixed $j$ (before imposing
the constraints (2.4)) are listed in Table 5.

We then proceeded as described above by diagonalizing $\hat D(n)^{2}$
separately on the superselection sectors corresponding to different
quantum numbers for the action of ${\cal O}^{(6)}\times\Z_{2}$. 
The resulting numbers of linearly independent
eigenstates of $\hat D(n)$ with strictly positive and vanishing
eigenvalues are listed in Tables 6 and 7.

\vskip1cm
{\offinterlineskip\tabskip=0pt
 \halign{ \strut\vrule#& \quad # \quad &\vrule#&
          \quad\hfil #\quad &
          \quad\hfil #\quad &
          \quad\hfil #\quad &
          \quad\hfil #\quad &
          \quad\hfil #\quad &
          \quad\hfil #\quad &
          \quad\hfil #\quad &
          \quad\hfil #\quad &
          \quad\hfil #\quad &
          \quad\hfil #\quad &\vrule#\cr \noalign{\hrule}
 & & & & & & & & & & && &\cr
 & \hfil j\hfil && 1 & 2 & 3 & 4 & 5 & 6 & 7 & 8 & 9 & 10 &\cr
 & & & & & & & & & & && &\cr
 \noalign{\hrule}
 & & & & & & & & & & && &\cr
 & $A_1$ &&
      1 & 2 & 5 & 8 & 14 & 20 & 30 & 40 & 55 & 70 &\cr
 & & & & & & & & & & && &\cr
 & $A_2$ &&
      0 & 0 & 0 & 1 & 2 & 5 & 8 & 14 & 20 & 30 &\cr
 & & & & & & & & & & && &\cr
 & $E$ &&
      0 & 2 & 6 & 14 & 26 & 44 & 68 & 100 & 140 & 190 &\cr
 & & & & & & & & & & && &\cr
 & total &&
      1 & 4 & 11 & 23 & 42 & 69 & 106 & 154 & 215 & 290 &\cr
 & & & & & & & & & & && &\cr
 \noalign{\hrule} }
\normalbaselines
\baselineskip=16pt   
\vskip.5cm
\line{{\bf Table 6} \hskip.3cm Numbers of linearly independent eigenstates of
$\hat D(n)$ with eigenvalue $>0$.\hfill} 
\vskip.7cm

\vskip.5cm
{\offinterlineskip\tabskip=0pt
 \halign{ \strut\vrule#& \quad # \quad &\vrule#&
          \quad\hfil #\quad &
          \quad\hfil #\quad &
          \quad\hfil #\quad &
          \quad\hfil #\quad &
          \quad\hfil #\quad &
          \quad\hfil #\quad &
          \quad\hfil #\quad &
          \quad\hfil #\quad &
          \quad\hfil #\quad &
          \quad\hfil #\quad &\vrule#\cr \noalign{\hrule}
 & & & & & & & & & & && &\cr
 & \hfil j\hfil && 1 & 2 & 3 & 4 & 5 & 6 & 7 & 8 & 9 & 10 &\cr
 & & & & & & & & & & && &\cr
 \noalign{\hrule}
 & & & & & & & & & & && &\cr
 & total &&
      3 & 7 & 12 & 19 & 27 & 37 & 48 & 61 & 75 & 91 &\cr
 & & & & & & & & & & && &\cr
 \noalign{\hrule} }
\normalbaselines
\baselineskip=16pt   
\vskip.5cm
\line{{\bf Table 7} \hskip.3cm Numbers of linearly independent eigenstates of
$\hat D(n)$ with eigenvalue $=0$.\hfill} 
\vskip.7cm

\vskip.5cm
{\offinterlineskip\tabskip=0pt
 \halign{ \strut\vrule#& \quad # \quad &\vrule#&
          \quad\hfil #\quad &
          \quad\hfil #\quad &
          \quad\hfil #\quad &
          \quad\hfil #\quad &
          \quad\hfil #\quad &
          \quad\hfil #\quad &
          \quad\hfil #\quad &
          \quad\hfil #\quad &
          \quad\hfil #\quad &
          \quad\hfil #\quad &\vrule#\cr \noalign{\hrule}
 & & & & & & & & & & && &\cr
 & \hfil j\hfil && 1 & 2 & 3 & 4 & 5 & 6 & 7 & 8 & 9 & 10 &\cr
 & & & & & & & & & & && &\cr
 \noalign{\hrule}
 & & & & & & & & & & && &\cr
 & total &&
      5 & 15 & 34 & 65 & 111 & 175 & 260 & 369 & 505 & 671 &\cr
 & & & & & & & & & & && &\cr
 \noalign{\hrule} }
\normalbaselines
\baselineskip=16pt   
\vskip.5cm
\line{{\bf Table 8} \hskip.3cm Number of linearly independent states
with fixed flux lines $\vec j=(j,j,j,j,j,j)$.\hfill} 
\vskip.7cm

\noindent Comparing the rows for the $A_{1}$-, $A_{2}$- and 
$E$-sectors in Table 6 to the total numbers of states in the
Hilbert spaces before the ${\cal O}^{(6)}\times T$-action is taken 
into account (Table 8), one observes that the matrix sizes are reduced 
considerably
when the super\-selection sectors are treated separately.
Another interesting feature is that the number of zero-eigenvalues
grows less rapidly with $j$ than that of the 
non-vanishing ones (by curve fits we found the dependences
$\frac{3}{4}j^{2}+\frac{3}{2}j+\frac{3}{4}$ for $j$ odd and
$\frac{3}{4}j^{2}+\frac{3}{2}j+1$ for $j$ even, as opposed to
$\frac{1}{2}j^{3}+\frac{3}{2}j^{2}+2j+1$).

Finally, here are our results for the non-negative eigenvalues of the operator
$\hat D(n)=\hat {\det E(n)}$. The numbers in Table 9 are given in 
length units, i.e. we have taken the sixth root of the eigenvalues 
of $\hat D(n)$. The three numbers in brackets give the degeneracy of 
eigenvalues in the $A_{1}$-, $A_{2}$- and $E$-sectors respectively.

\vskip.7cm
{\offinterlineskip\tabskip=0pt
 \halign{ \strut\vrule#& \quad # \quad &\vrule#&
          \quad #\hfil \quad &\vrule#\cr \noalign{\hrule}
 & && &\cr
 & $j=1$ && 0.820977 (1,0,0), 0 (3) &\cr
 & && &\cr
 & $j=2$ && 1.07707 (1,0,0), 0.933429 (1,0,2), 0 (7) & \cr
 & && &\cr
 & $j=3$ && 1.28653 (1,0,0), 1.17666 (1,0,2), 1.03890 (1,0,2), 
            1.02292 (1,0,2), & \cr
 & && 0.915413 (1,0,0), 0 (12) & \cr
 & && &\cr
 & $j=4$ && 1.46726 (1,0,0), 1.37521 (1,0,2), 1.27163 (1,0,2), 
            1.25951 (1,0,2), & \cr 
 & && 1.16697 (1,0,0), 1.12099 (1,1,4), 1.09850 (1,0,2), 
      1.01217 (1,0,2), 0 (19) & \cr
 & && &\cr
 & $j=5$ && 1.62835 (1,0,0), 1.54767 (1,0,2), 1.46030 (1,0,2), 
      1.45180 (1,0,2), & \cr
 & && 1.37044 (1,0,0), 1.35016 (1,1,4), 1.33173 (1,0,2), 
      1.25452 (1,0,2), & \cr 
 & && 1.19678 (1,0,2), 1.19140 (1,1,4), 1.16425 (1,0,2), 
      1.10871 (1,0,2), & \cr
 & && 1.09057 (1,0,2), 0.978018 (1,0,0), 0 (27) & \cr
 & && &\cr
 & $j=6$ && 1.77500 (1,0,0), 1.70234 (1,0,2), 1.62518 (1,0,2), 
      1.61898 (1,0,2), & \cr
 & && 1.54569 (1,0,0), 1.53375 (1,1,4), 1.52004 (1,0,2), 
      1.44981 (1,0,2), & \cr 
 & && 1.42448 (1,0,2), 1.41907 (1,1,4), 1.39619 (1,0,2), 
      1.34168 (1,0,2), & \cr
 & && 1.32815 (1,0,2), 1.26214 (1,1,4), 1.25364 (1,1,4), 
      1.23241 (1,0,0), & \cr
 & && 1.22270 (1,0,2), 1.18305 (1,1,4), 1.15815 (1,0,2), 
      1.06919 (1,0,2), 0 (37) & \cr
 & && &\cr
 & $j=7$ &&  1.91050 (1,0,0), 1.84387 (1,0,2), 1.77395 (1,0,2), 
      1.76921 (1,0,2), & \cr 
 & && 1.70204 (1,0,0), 1.69398 (1,1,4), 1.68357 (1,0,2), 
      1.61870 (1,0,2), & \cr
 & && 1.60414 (1,0,2), 1.59947 (1,1,4), 1.58195 (1,0,2), 
      1.52832 (1,0,2), & \cr
 & && 1.51891 (1,0,2), 1.48980 (1,1,4), 1.48107 (1,1,4), 
      1.45463 (1,0,2), & \cr
 & && 1.43533 (1,0,0), 1.41349 (1,1,4), 1.39331 (1,0,2), 
      1.32314 (1,0,2), & \cr
 & && 1.32056 (1,1,4), 1.31437 (1,0,2), 1.30965 (1,1,4), 
      1.27555 (1,0,2), & \cr
 & && 1.25331 (1,0,2), 1.24732 (1,1,4), 1.21785 (1,0,2), 
      1.16207 (1,0,2), & \cr
 & &&1.14257 (1,0,2), 1.02552 (1,0,0), 0 (48)& \cr
 & && &\cr
 & $j=8$ &&  2.03705 (1,0,0), 1.97517 (1,0,2), 1.91074 (1,0,2), 
      1.90697 (1,0,2), & \cr
 & && 1.84463 (1,0,0), 1.83874 (1,1,4), 1.83056 (1,0,2), 
      1.77001 (1,0,2), & \cr
 & && 1.76040 (1,0,2), 1.75651 (1,1,4), 1.74290 (1,0,2), 
      1.69067 (1,0,2), & \cr
 & && 1.68385 (1,0,2), 1.66720 (1,1,4), 1.65942 (1,1,4), 
      1.63881 (1,0,2), & \cr
 & && 1.60893 (1,0,0), 1.59606 (1,1,4), 1.58113 (1,0,2), 
      1.55187 (1,0,2), & \cr
 & && 1.54891 (1,1,4), 1.53771 (1,1,4), 1.50965 (1,0,2), 
      1.50826 (1,0,2), & \cr
 & && 1.48282 (1,0,2), 1.47687 (1,1,4), 1.45221 (1,0,2), 
      1.39757 (1,0,2), & \cr
 & && 1.38309 (1,0,2), 1.37801 (1,1,4), 1.37367 (1,1,4), 
      1.36072 (1,1,4), & \cr
 & && 1.32393 (1,0,2), 1.31397 (1,1,4), 1.30466 (1,1,4), 
      1.28429 (1,0,0), & \cr
 & && 1.27155 (1,0,2), 1.23247 (1,1,4), 1.20595 (1,0,2), 
      1.11432 (1,0,2), 0 (61)& \cr
 & && &\cr
 & $j=9$ && 2.15620 (1,0,0), 2.09818 (1,0,2), 2.03812 (1,0,2), 
      2.03504 (1,0,2), & \cr
 & && 1.97663 (1,0,0), 1.97209 (1,1,4), 1.96545 (1,0,2), 
      1.90848 (1,0,2), & \cr
 & && 1.90159 (1,0,2), 1.89833 (1,1,4), 1.88744 (1,0,2), 
      1.83691 (1,0,2), & \cr
 & && 1.83172 (1,0,2), 1.82081 (1,1,4), 1.81415 (1,1,4), 
      1.79792 (1,0,2), & \cr
 & && 1.76329 (1,0,0), 1.75461 (1,1,4), 1.74332 (1,0,2), 
      1.72777 (1,0,2), & \cr
 & && 1.72491 (1,1,4), 1.71478 (1,1,4), 1.69154 (1,0,2), 
      1.67749 (1,0,2), & \cr
 & && 1.66205 (1,0,2), 1.65696 (1,1,4), 1.63812 (1,0,2), 
      1.60819 (1,1,4), & \cr
 & && 1.60323 (1,1,4), 1.58998 (1,1,4), 1.58435 (1,0,2), 
      1.57430 (1,0,2), & \cr
 & && 1.55793 (1,0,2), 1.54379 (1,1,4), 1.53433 (1,1,4), 
      1.50616 (1,0,2), & \cr
 & && 1.48851 (1,0,0), 1.46554 (1,1,4), 1.44411 (1,0,2), 
      1.42977 (1,0,2), & \cr
 & && 1.42829 (1,1,4), 1.42251 (1,1,4), 1.40779 (1,1,4), 
      1.37128 (1,0,2), & \cr
 & && 1.36866 (1,0,2), 1.36845 (1,1,4), 1.36330 (1,0,2), 
      1.35667 (1,1,4), & \cr
 & && 1.32054 (1,0,2), 1.29967 (1,0,2), 1.29326 (1,1,4), 
      1.26211 (1,0,2), & \cr
 & && 1.20555 (1,0,2), 1.18508 (1,0,2), 1.06412 (1,0,0), 0 (75) & \cr
 & && &\cr
 & $j=10$ && 2.26912 (1,0,0), 2.21432 (1,0,2), 2.15784 (1,0,2), 
      2.15526 (1,0,2), & \cr
 & && 2.10012 (1,0,0), 2.09649 (1,1,4), 2.09097 (1,0,2), 
      2.03703 (1,0,2), & \cr
 & && 2.03180 (1,0,2), 2.02905 (1,1,4), 2.02009 (1,0,2), 
      1.97130 (1,0,2), & \cr
 & && 1.96720 (1,0,2), 1.95944 (1,1,4), 1.95376 (1,1,4), 
      1.94061 (1,0,2), & \cr
 & && 1.90385 (1,0,0), 1.89752 (1,1,4), 1.88867 (1,0,2), 
      1.87925 (1,0,2), & \cr
 & && 1.87664 (1,1,4), 1.86783 (1,1,4), 1.84934 (1,0,2), 
      1.82734 (1,0,2), & \cr
 & && 1.81716 (1,0,2), 1.81294 (1,1,4), 1.79834 (1,0,2), 
      1.78328 (1,1,4), & \cr
 & && 1.77840 (1,1,4), 1.76633 (1,1,4), 1.74596 (1,0,2), 
      1.74079 (1,0,2), & \cr
 & && 1.73868 (1,0,2), 1.72121 (1,1,4), 1.71284 (1,1,4), 
      1.69092 (1,0,2), & \cr
 & && 1.66208 (1,0,0), 1.66200 (1,0,2), 1.66017 (1,1,4), 
      1.65364 (1,1,4), & \cr
 & && 1.64853 (1,1,4), 1.63864 (1,1,4), 1.63269 (1,0,2), 
      1.60428 (1,0,2), & \cr
 & && 1.60238 (1,0,2), 1.59918 (1,1,4), 1.58717 (1,1,4), 
      1.55981 (1,0,2), & \cr
 & && 1.55607 (1,0,2), 1.53183 (1,0,2), 1.52550 (1,1,4), 
      1.49950 (1,0,2), & \cr
 & && 1.47744 (1,1,4), 1.47486 (1,1,4), 1.46780 (1,1,4), 
      1.45154 (1,1,4), & \cr
 & && 1.44430 (1,0,2), 1.42912 (1,0,2), 1.42294 (1,1,4), 
      1.41825 (1,1,4), & \cr
 & && 1.41035 (1,0,2), 1.40442 (1,1,4), 1.36572 (1,0,2), 
      1.35747 (1,1,4), & \cr
 & && 1.34758 (1,1,4), 1.32752 (1,0,0), 1.31283 (1,0,2), 
      1.27374 (1,1,4), & \cr
 & && 1.24600 (1,0,2), 1.15189 (1,0,2), 0 (91) & \cr
 & && &\cr
\noalign{\hrule} }
\normalbaselines
\baselineskip=16pt   
\vskip.5cm
\line{{\bf Table 9} \hskip.3cm Values for $x^{\frac{1}{6}}$, where $x$
is a non-negative eigenvalue of $\hat D(n)$.\hfill}
\vskip.7cm

The spectrum is rather complex, and becomes more and more spread out
with increasing $j$. There seem to be only three different degeneracy
patterns, $(1,0,0)$, $(1,0,2)$ and $(1,1,4)$. 
Every eigenvalue that occurs is already contained in the 
${\cal O}^{(6)}$-invariant (i.e. the $A_{1}$-) sector, and is 
non-degenerate. It would be interesting to see whether this is 
also the case for more general flux line configurations. Note also 
that the highest eigenvalue for fixed $j$ is always non-degenerate. 
The ``volume deficit'' observed in [24] (the fact that all
volume eigenvalues are systematically smaller than expected 
from the (Laplacian) area eigenvalues, compared to the relation
one would obtain for a Euclidean reference metric), persists for
higher $j$, although it becomes less pronounced. 

\vskip1.5cm
\line{\ch 4 Summary and outlook\hfil}

We have explained in this paper how the diagonalization of the
volume operator in the loop representation of quantum gravity
can be simplified by taking into account discrete symmetries at
the intersections of the Wilson loop states. 
Applied to the case of
up to six-valent intersections on a cubic lattice, this requires
the decomposition of Hilbert space into the irreducible
representations of a six-dimensional subgroup ${\cal O}^{(6)}$ 
of the cubic group.
One finds a set of local operators that commute with $\hat {\det E}$,
and therefore can be diagonalized simultaneously.
A further simplification arises when one includes the total spatial
reflection, and diagonalizes the operator $(\hat {\det E})^2$, from
which the eigenstates of $\hat {\det E}$ can be obtained using the
results in [17]. 
To demonstrate the viability of the method, we have calculated the
spectrum for a class of six-valent intersections with flux line
numbers $(j,j,j,j,j,j)$, for $j\leq 10$. 
We found that all eigenvalues are already contained in the 
${\cal O}^{(6)}$-invariant sector, without degeneracy. 
It is conceivable that this sector is also distinguished on
physical grounds, but this depends on how the continuum limit 
of the lattice theory will be taken. 

Let us close with some comments on the condition $\hat {\det E}>0$ which,
as we have argued in [17], should be imposed on the quantum state space.
It remains to be seen at which stage of the quantization it is imposed most
conveniently. Ignoring for the moment the zero-eigenvalue states,
this condition reduces the dimensionality of the Hilbert space by
a factor $2^x$, where $x$ is the number of intersections of the lattice.
In addition, one obtains a condition on physical operators $\hat{\cal P}$, 
namely, that they should not map out of the subspace of states with
$\hat {\det E}$-eigenvalues $\geq 0$ (or $>0$). A sufficient condition is
given by the vanishing of the commutator

$$
[\hat {\cal P},{\rm sign}(\hat {\det E})]=0.\eqno(4.1)
$$

However, we have not found
a simple form of this condition which would not require the 
explicit knowledge of the eigenstates of $\hat {\det E}$. 
Note that (4.1) is a rather strong condition which, for example, is
not satisfied by the lattice analogues 
of the area operators defined in [25]. A less stringent condition
is to require that (4.1) be satisfied in the limit as the lattice
spacing is taken to zero. This is also sensible from a physical point of
view, since the condition should be independent of the regularization
(for example, the version of the area operator corresponding
to a pure Laplacian [10,20] obviously fulfills (4.1), but coincides with
other discretized forms of the area to lowest order in the lattice
spacing). 

We do not know whether there is a way to formulate a condition like 
$\hat {\det E}>0$ in the continuum theory. The regularization
used for the volume operator in [14] forces one to use modulus signs
around the operator, in order that the square root
$\sqrt{|\hat {\det E}|}$ is well-defined, something not
necessary in the case of the lattice regularization. 
Nevertheless, we think that such a constraint on states,
along with operator conditions of the type (4.1),
should be imposed in the quantum theory --
even if this leads to new complications -- 
because they describe a property of the gravitational theory.

\vskip2cm
\line{\ch References\hfil}

\item{[1]} A. Ashtekar: New variables for classical and quantum
  gravity, {\it Phys. Rev. Lett.} 57 (1986) 2244-7; A new 
  Hamiltonian formulation of general relativity, {\it Phys.
  Rev.} D36 (1987) 1587-1603.

\item{[2]} R.S. Tate: Polynomial constraints for general relativity
  using real geometrodynamical variables, {\it Class. Quant.
  Grav.} 9 (1992) 101-19.

\item{[3]} C. Rovelli and L. Smolin: Loop space representation of
  quantum general relativity, {\it Nucl. Phys.} B331 (1990) 80-152.

\item{[4]} articles by B. Br\"ugmann, R. Loll, T. Thiemann and
  A. Ashtekar {\it in} Canonical gravity: from classical to quantum,
  eds. J. Ehlers and H. Friedrich, Springer, Berlin, 1994;
  R. Loll: Still on the way to quantizing gravity, {\it preprint}
  Potsdam AEI-025, 1997.
 
\item{[5]} P. Renteln and L. Smolin: A lattice approach to spinorial
  quantum gravity, {\it Class. Quant. Grav.} 6 (1989) 275-94;
  P. Renteln: Some results of SU(2) spinorial lattice
  gravity, {\it Class. Quant. Grav.} 7 (1990) 493-502;
  O. Bostr\"om, M. Miller and L. Smolin: A new discretization of
  classical and quantum general relativity, Syracuse U. {\it preprint} 
  SU-GP-93-4-1;
  R. Loll: Non-perturbative solutions for lattice quantum gravity,
  {\it Nucl. Phys.} B444 (1995) 619-39;
  K. Ezawa: Multi-plaquette solutions for discretized 
  Ashtekar gravity, {\it Mod. Phys. Lett.} A11 (1996) 2921-32;
  H. Fort, R. Gambini and J. Pullin: Lattice knot theory and quantum
  gravity in the loop representation, Penn State U. {\it preprint}
  CGPG-96/8-1.

\item{[6]} J.F. Barbero G.: Real Ashtekar variables for Lorentzian
  signature space-times, {\it Phys. Rev.} D51 (1995) 5507-10.

\item{[7]} R. Loll: The volume operator in discretized quantum
  gravity, {\it Phys. Rev. Lett.} 75 (1995) 3048-51.

\item{[8]} R. Loll: Spectrum of the volume operator in
  quantum gravity, {\it Nucl. Phys.} B460 (1996) 143-54.

\item{[9]} L. Smolin: Recent developments in nonperturbative quantum
  gravity, {\it in} Quantum gravity and cosmology, World Scientific,
  Singapore, 1992, 3-84.

\item{[10]} C. Rovelli and L. Smolin: Discreteness of area and 
  volume in quantum gravity, {\it Nucl. Phys.} B442 (1995)
  593-622, Err. {\it ibid.} B456 (1995) 753-4.

\item{[11]} R. De Pietri and C. Rovelli: Geometry eigenvalues
  and scalar product from recoupling theory in loop quantum gravity,
  {\it Phys. Rev.} D54 (1996) 2664-90.

\item{[12]} A. Ashtekar and J. Lewandowski: Differential geometry on the 
  spaces of connections via graphs and projective limits, 
  {\it J. Geom. Phys.} 17 (1995) 191-230.

\item{[13]} J. Lewandowski: Volume and quantizations, {\it Class.
  Quant. Grav.} 15 (1997) 71-6.

\item{[14]} T. Thiemann: Closed formula for the matrix elements of the volume
  operator in canonical quantum gravity, Harvard U. {\it preprint}
  HUTMP-96/B-353.

\item{[15]} R. Loll: A real alternative to quantum gravity in loop
  space, {\it Phys. Rev.} D54 (1996) 5381-4.

\item{[16]} T. Thiemann: Anomaly-free formulation of nonperturbative 
  four-dimensional Lorentzian quantum gravity, {\it Phys. Lett.} B380
  (1996) 257-64; Quantum spin dynamics I\&II, Harvard U. 
  {\it preprints} HUTMP-96/B-351 and B-352.

\item{[17]} R. Loll: Imposing ${\det E}>0$ in discrete quantum
  gravity, {\it Phys. Lett.} B399 (1997) 227-32.

\item{[18]} L. Smolin: Instability and absence of long-ranged
  correlations in non-perturbative quantum general relativity,
  Penn State U. {\it preprint} CGPG-96/9-4.

\item{[19]} C. Rovelli and L. Smolin: Spin networks and quantum 
  gravity, {\it Phys. Rev.} D52 (1995) 5743-59;
  J.B. Baez: Spin network states in gauge theory, to
  appear in {\it Adv. Math.};
  Spin networks and nonperturbative quantum gravity, 
  UC Riverside {\it preprint}, 1995; L. Smolin: The future of
  spin networks, Penn State U. {\it preprint} CGPG-97.

\item{[20]} R. Loll: Further results on geometric operators in quantum
  gravity, to appear in {\it Class. Quant. Grav.}
 
\item{[21]} T. Janssen: Crystallographic groups, North-Holland, 1973;
  S.L. Altmann, {\it in} Quantum Theory II, ed.
  D.R. Bates, Academic Press, 1962.

\item{[22]} J. Kogut and L. Susskind: Hamiltonian formulation of
  Wilson's lattice gauge theories, {\it Phys. Rev.} D11 (1975)
  395-408; J.B. Kogut: The lattice gauge theory approach
  to quantum chromodynamics, {\it Rev. Mod. Phys.} 55 (1983) 775-836.

\item{[23]} B. Berg and A. Billoire: Glueball spectroscopy in 4d SU(3)
  lattice gauge theory(I), {\it Nucl. Phys.} B221 (1983) 109-40.
  
\item{[24]} R. Loll: Latticing quantum gravity, Potsdam {\it preprint}
  AEI-024, 1997.

\item{[25]} A. Ashtekar and J. Lewandowski: Quantum theory of geometry I:
  area operators, {\it Class. Quant. Grav.} 14 (1997) A55-82.

\end